\documentclass[12pt]{article}
\usepackage{epsfig}
\usepackage{cite}
\usepackage{./mcite}
\usepackage{url}

\chardef\til=126

\newcommand{\gev}{{\,\mathrm{GeV}}}

\begin{document}

\clearpage
\pagestyle{empty}
\setcounter{footnote}{0}\setcounter{page}{0}%
\thispagestyle{empty}\pagestyle{plain}\pagenumbering{arabic}%

\hfill ANL-HEP-PR-05-118
 
\hfill November  2005

\hfill Updated June 2006

\vspace{2.0cm}

\begin{center}

{\Large\bf
A new jet algorithm based on the $k$-means clustering
for the reconstruction of heavy states from jets
\\[-1cm] }

\vspace{2.5cm}

{\large S.~Chekanov\footnote[1]{Also affiliated with DESY, Notkestrasse 85, 22607, Hamburg, Germany} }

\vspace{0.5cm}

HEP division, Argonne National Laboratory, \\
9700 S.Cass Avenue,
Argonne, IL 60439
USA \\
E-mail: chekanov@mail.desy.de

\vspace{0.5cm}
\begin{abstract}
A jet algorithm based on the $k$-means clustering procedure 
is proposed which can be used for the invariant-mass reconstruction of
heavy states decaying to hadronic jets. 
The proposed algorithm was tested by reconstructing
$e^{+}e^{-}\to
t\overline{t} \to 6\;\hbox{\rm jets}$ and
$e^+e^- \to W^+W^- \to 4 \;\hbox{\rm jets}$
processes  at $\sqrt{s}=500\gev$ using a Monte Carlo simulation.
It was shown that the algorithm
has a reconstruction efficiency similar to 
traditional jet-finding algorithms,     
and leads to 
$25\%$ and $40\%$ improvement of the top-quark and
$W$ mass resolution, respectively, compared  to 
the $k_T$ (Durham) algorithm.
In addition, it is expected that the peak
positions measured with the new algorithm
have smaller systematical uncertainty.
\end{abstract}

\end{center}

\newpage
\setcounter{page}{1}

\section{Introduction}

Jet finding algorithms are  indispensable
tools for the reconstruction of heavy states ($Z,W$ bosons, top quarks, Higgs bosons) 
decaying to hadronic jets.
A  number of jet algorithms has been proposed in the past (see recent
reviews \cite{Moretti:1998qx,chekk}) which can be used  
for the calculation of  the invariant-mass distributions for hadronically
decaying heavy states.

It has already been pointed out \cite{Moretti:1998qx} that there is no 
algorithm which is optimal for all 
possible jet-related studies. Usually, different jet algorithms have different emphasis.
Some jet finders are preferable for precise comparisons with QCD
theory, since the jet cross sections reconstructed with such algorithms have small 
fixed-order perturbative corrections, as well as  small hadronisation corrections.  
However, such jet algorithms may not be the most optimal for other tasks.

The traditional jet finders have one significant drawback:
miss-assignment of hadrons into jets is a common problem  for the reconstruction
of heavy states decaying into jets. Incorrectly assigned particles lead
to a broadening of the width of the invariant-mass peaks, as well as 
to a reduction of signal-over-background ratios. To deal with this
problem, one can impose expected  kinematic criteria on 
the reconstructed jets.
However, the construction of the traditional algorithms  prevents to 
include such criteria in an efficient way:  
the iterative procedure which combines particles into jets
is usually based on a single distance measure between particles. 
Therefore,  it is difficult to take into account
{\em a priory} known information on decay kinematics 
during the jet clustering procedure.

To solve the miss-assignment problem, 
one may think about an iterative procedure which would keep
redistributing  hadrons
between jets until known kinematic criteria are met.
In this case,  the main question is how the  particles
should be redistributed (particles in jets with the strongest overlaps?) and what
``particle-redistribution  algorithm''  should be used for this, keeping in mind that the speed for 
such procedure should be reasonably fast.

Below we will discuss an algorithm which  attempts  
to solve the problem of particle miss-assignments.
In fact, we propose a jet clustering procedure
with some additional elements of intelligence: it minimises not only a distance
measure between hadrons, but also any physics-related  quantity reflecting 
how close the final event kinematics is from the expected one.
To illustrate its properties, we will consider
$e^{+}e^{-}\to
t\overline{t}\to  b\bar{b} W^{+}W^{-} \to  6\;\hbox{\rm jets}$
and $e^{+}e^{-}\to
W^{+}W^{-}\to 4\;\hbox{\rm jets}$
decays at $\sqrt{s}=500\gev$.
We have chosen such processes due to their simplicity, since 
the event signatures are characterised by the production of
exactly six (four)  hadronic jets. The all-hadronic top decay is also  considered to be 
the most promising for top studies at the International  Linear Collider (ILC), 
since this channel has the largest branching ratio 
($\simeq 44\%$ of all $t\overline{t}$ decays).

\section{$k$-means clustering algorithm}

We will remind that
the $k$-means \cite{kmeans} clustering 
is among the oldest (and simplest) unsupervised learning algorithms
that solve clustering problems.  It has been adapted  to classify the data 
in many problem domains. Below we will remind of the $k$-means procedure.

Let us assume that we have $N$ particles  and we know
that all these particles should be grouped 
to a fixed number $N_{cl}$ of clusters. 
The main idea is to define the locations for the 
initial $N_{cl}$  centroids, or center points, in a certain
phase space.
These centroids should be placed
as much as possible far away from each other.
The next step is to associate each point belonging to a given data set 
to the nearest centroid. In the simplest approach, 
one could use  a minimum-distance classifier to assign all particles
to such centroids. Once this assignment is done, then 
the positions of new  centroids should be recalculated.
This procedure is repeated in a loop.  
As a result of such iteration,  the  centroids
change their location step by step until they do not move any more.
For the final cluster configuration,  
each data point will be associated to the closest centroid. 

The grouping is usually 
done by minimising the sum of squares of the distances between  data points 
and the corresponding cluster centroid,
although other choices are also possible.
For this simplest choice of the metrics, the algorithm minimises the quantity:
\begin{equation}
S=\sum_{k=1}^{N_{cl}}\sum_{n \in L_k} \mid x_n-C_k\mid^2,
\label{eq1}
\end{equation}
where $x_n$ is a vector representing the $n^{th}$ data point  and $C_k$ is
the geometric location of the cluster center in the subset $L_k$ (i.e.  
the data points associated with the $k$th cluster centroid). 
It  can be proved that the $k$-means procedure always terminates for this metrics. 
However, the $k$-means algorithm
does not necessarily find the most optimal configuration, 
and it has a  significant sensitivity to the
initial, randomly selected, centroid locations. Thus the algorithm should be
run multiple times to reduce this instability effect.

The last feature  could  help to construct an ``intelligent'' 
algorithm which minimises
not only a distance measure between particles and the
centroids (i.e. jet centers), but also any physics-related optimisation criteria. 
To be more specific, let us consider an example which is relevant 
for high-energy physics: $e^{+}e^{-}\to t\overline{t}\to
b\overline{b} W^{+}W^{-}\to 6\;\hbox{\rm jets}$ process.
In accordance with the topology of such events, we should expect that
all hadrons should be clustered into six jets. 
Thus, six centroids (i.e. jet  seeds) randomly located in a phase space
should be specified for the initial $k$-means clustering loop.
The clustering can be  performed by minimizing
the distances from the centroids to hadrons in the azimuthal angle ($\phi$)
and rapidity ($y$) phase space.  
After the end of the initial iterative procedure, the cluster topology
can be characterised by the sum $S$ of the distances from the centers of the 
jets to hadrons,  as given by Eq.~(\ref{eq1}).
The procedure should be repeated $K$ times using
different starting locations for the centroids.
This gives $K$ solutions with the final values of the
metrics $S_1,\ldots S_K$. 
The number $K$ should be large enough to make sure that 
there are several configurations with the same $S_i$. 
This leads to a confidence that all possible configurations were
explored and that an absolute minimum can be found. 
If there are several final configurations with 
the smallest $S_i$ (which are exactly the same),  
then one could say that a hadron assignment with the strongest
particle collimation inside jets is found. It can be characterised by
$S_{\mathrm{min}}$.

Note that the final configuration is the most optimal 
from the point of view of  closeness of hadrons to the central jet positions.
Certainly, it may not be the most optimal from the physics point of view since
some hadrons  (located mostly at the edge of the jets) could still be assigned to
wrong jets. To minimise this problem, one can use kinematic
requirements already during the $k-$means clustering iterations.
In order to take into account known event kinematics, one could multiply
$S_i$ by a weight factor which can reflect a likeliness
of a certain cluster configuration from the point of view of
the expected physics output. The  weight factor
can be proportional to  $\sim 1- P_i$,
where $P_i$ is the probability of how close  a particular cluster configuration is to
the expected one.  For example, for the fully-hadronic $t\bar{t}$ production, 
$S_i$ should be reduced if there are at least two dijets in an event with the invariant masses close
to the $W$-boson mass.

The traditional jet finders only minimise a certain distance measure between particles.
For such jet algorithms, once the particle assignment is done,
the event could  either be taken (if, for example, there are two 
jets with the masses close to the $W$ for the all-hadronic top decays) 
or rejected (in the opposite case).
Thus, the event-kinematic requirements are completely external and independent 
of the jet finding procedure. 
In contrast, such requirements  are an essential part of the
proposed jet clustering. This means that the
new algorithm  keeps analysing the same event by trying different
final configurations until certain kinematics conditions are satisfied. 
Events can only be rejected if
it is not possible to find such an assignment of hadrons which meets
the criterion of the closeness of hadrons to jet centers
and at the same time satisfies expected physics requirements.

For a single event,  the $k$-means minimisation procedure
leads to different locations of the jet centers, as
well as to different assignment of particles into the jets. Typically, the particle 
assignments with different initial seeds are not drastically different one from the other. 
Therefore, one could view the overall picture as a redistribution of hadrons (mainly located
in the regions of  strongest jet overlaps)  between the jets with fixed centers for all
$k$-means  configurations which differ one from the other by different 
initial conditions.

If the produced jets are very well collimated, then one should expect a small difference
between the proposed $k$-means clustering and the standard jet finding algorithms:  
in this case all $k$-means cluster configurations with different initial
centroids should give identical results (i.e. all $S_i$ will be the same).
In contrast, the constrained $k-$means algorithm could outperform the standard algorithms
for events with broad and overlapping jets.   

\section{Top-quark production}

\subsection{Durham jet finder versus unconstrained  
            $k$-means \\ clustering algorithm}

To illustrate the method outlined above, we will apply it to the
all-hadronic top decays in $e^+e^-$ annihilation
at the centre-of-mass energy of $\sqrt{s}=500\gev$.
The  PYTHIA 6.3 model~\cite{cpc:39:347,*Sjostrand:2003wg}
was used to  generate one million of fully
inclusive $e^+e^-$ events, including the $t\bar{t}$  production.
This sample contains 14740 events with fully-hadronic top decays.   
The default PYTHIA parameters were used for the simulation.
The initial-state photon radiation was included.
The mass and the Breit-Wigner width of the top quarks were set to
the defaults values, $175\gev$ and $1.39\gev$, respectively.
The particles with the lifetime more than 3 cm were
considered to be stable. Neutrinos were removed from the consideration.
We require all reconstructed jets to have the energies above $10\gev$.
In order to remove events with a large fraction of neutrinos, we apply
the momentum and the energy imbalance cuts  
similar to those used in \cite{chekmorg}:
\begin{equation}
\mid \frac{E_{vis}}{\sqrt{s}}-1\mid <0.07,
\quad \frac{\mid \sum \vec{p}_{||i}\mid }{\sum \mid \overrightarrow{p_{i}}\mid }<0.04,
\quad \frac{\mid \sum  \vec{p}_{Ti}  \mid }
{\sum \mid \overrightarrow{p_{i}}\mid }<0.04,
\label{eq:1}
\end{equation}
where $E_{vis}$ is the visible energy, $\vec{p}_{||i}$ ($\vec{p}_{Ti}$) is
the longitudinal (transverse) component of momentum of a final-state particle and 
the sum runs over all final-state particles.

We do not use a detector simulation for the generated events 
since  such study is outside of the scope of this paper. Here we 
address the issue of the reconstruction of the invariant masses
which are smeared with respect to the true masses 
by the parton shower and hadronisation effects. 
Also, for simplicity, no  $b-$tagging requirement was assumed.

\begin{figure}
\begin{center}
\mbox{\epsfig{file=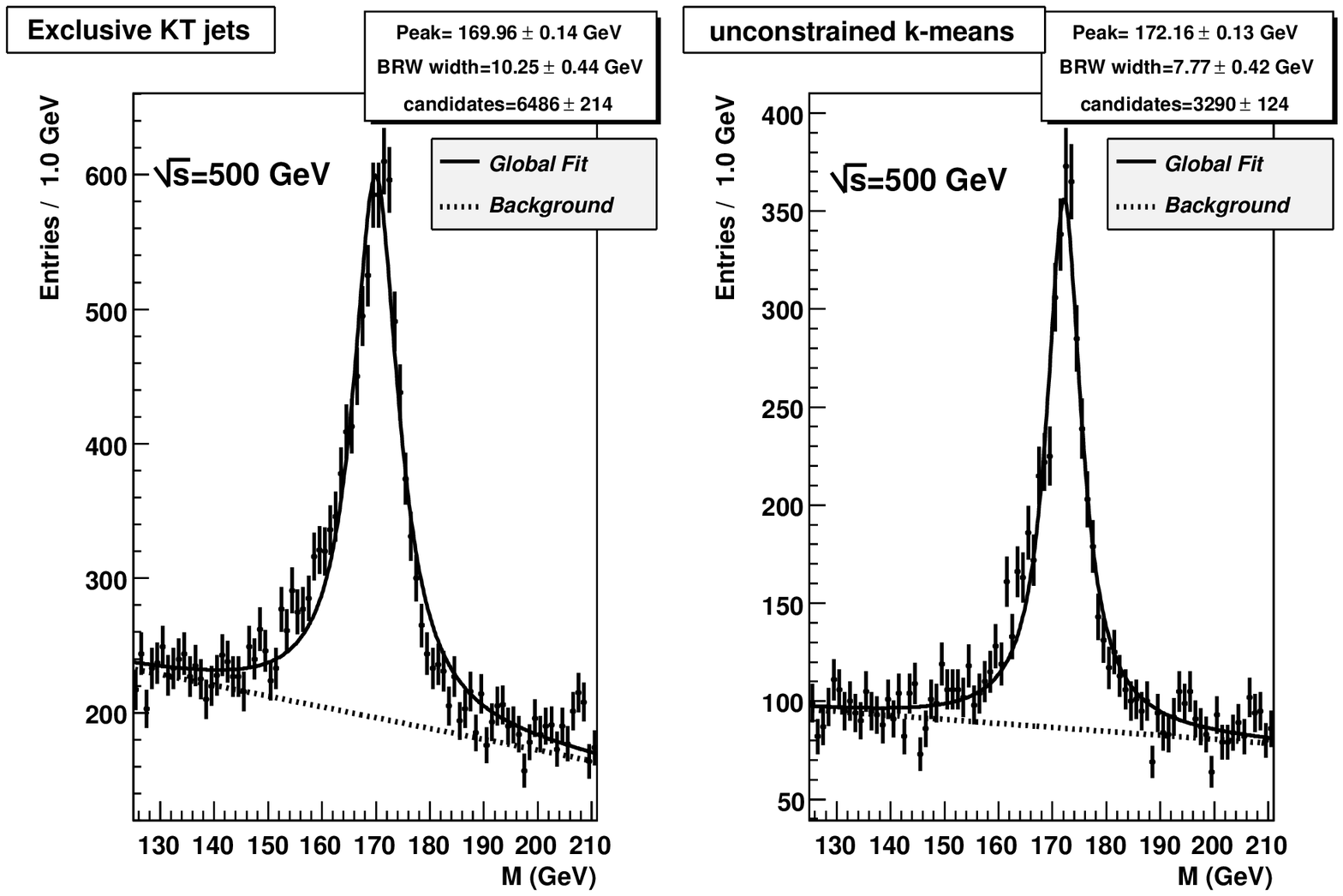,height=10cm}}
\caption
{
The distribution of the trijet invariant masses for the reconstruction of 
all-hadronic top decays. Fully inclusive $e^+e^-$ events were 
generated with PYTHIA for $\sqrt{s}=500\gev$.
The reconstruction was done using the $k_T$ algorithm (left) and the $k$-means 
algorithm (right).   
The fit was performed using the  Breit-Wigner function together with a second-order
polynomial to describe the background.
}
\label{ttfig1}
\end{center}
\end{figure}

First, the reconstruction was done using the traditional method:
jets where found  using the exclusive mode of the
$k_{\bot }$ (Durham) algorithm \cite{pl:b269:432}, requiring exactly six jets
for each event. 
Our choice for the Durham  algorithm was motivated by the fact that 
this jet finder is one of the
best algorithms for the reconstruction of jet invariant masses in $e^+e^-$, as
it was illustrated using  the $W$-mass reconstruction example \cite{Moretti:1998qx}.
We use a C++ version of this jet algorithm~\cite{Butterworth:2002xg}.
The event is taken if there is at least one jet-pair with the invariant mass $M_{jj}$
in the range $M_W \pm 10\gev$, where $M_W$ is the nominal mass of the $W$ boson.
Next, the dijets which passed this cut were combined with the rest of the jets,
and then all three-jet combinations were plotted.
Figure~\ref{ttfig1}(left) shows the corresponding trijet invariant masses, $M_{jjj}$. 
The fit was performed using the Breit-Wigner function together with a second-order 
polynomial for the background description.
The  reconstructed Breit-Wigner width ($\simeq 10$ GeV) is similar
to that when an alternative approach for the top reconstruction
was used \cite{chekmorg}. The method discussed  in Ref.~\cite{chekmorg} 
does not use the assumption on the $W$ mass. 

Now let us consider the $k$-means algorithm.  As a first step, 
the final-state hadrons were pre-clustered
with the Durham algorithm using $y^{cut}_{\mathrm{min}} = 10^{-5}$.
This procedure reduces the number of data points by a factor 3--6.
The average number of the final subjets for the 
$t\bar{t}$ production was around 20.
As it will be discussed below, 
this step was necessary  to reduce the computational time. The $k$ means
algorithm was run on the subjets.
Each $e^+e^-$ event was analysed $K=300$ times, every time using different (random)
locations for the initial centroids. 
This number was found to be sufficiently large to explore
all possible jet configurations.

The subjet clustering was performed in the rapidity and the azimuthal angle.
For the $k-$means clustering, it is commonly accepted 
to normalize each variable by its standard deviation. 
Therefore,   
both variables were normalized such that their 
available range was approximately between 0 and 1. 
Without such transformation, 
the number of the reconstructed
states to be discussed below is $5-8\%$ lower than
in case when the transformation is used.

After the $k$-means clustering, each $e^+e^-$ event is characterised
by the set $S_i,\,  i=1,\ldots K$,   where $S_i$ denotes the sum 
of all distances from the centers of the $k$-means jets to hadrons.
Only jet configurations with the same smallest $S_{i}$ were  accepted. 
Typically, there are 10--20 final configurations 
which are characterised by the same $S_{\mathrm{min}}$.
The result of the $M_{jjj}$ reconstruction 
is shown in Fig.~\ref{ttfig1}(right). 
The  $M_{jjj}$ masses were plotted only for configurations 
characterised by the minimum $S_{\mathrm{min}}$.
It can be seen that the $k$-means 
algorithm leads to a better mass resolution (width) than 
the  Durham jet finder.  In addition, the reconstructed 
peak position is closer 
to the generated top mass ($175$ GeV).  An obvious drawback of 
the standard $k$-means algorithm is a smaller reconstruction efficiency (i.e. a smaller
number of the reconstructed events) than for the Durham algorithm, since 
the $k$-means algorithm in its present form has a tendency to produce 
low-energy jets ($<10\gev$).  
Below we will discuss how to improve the $k$-means procedure.

\subsection{Constrained $k$-means algorithm}

Let us again consider the $k$-means algorithm, but this time we will
constrain it by some requirement:   
each $S_i$  will be multiplied by an additional weight factor.
This factor is constructed from several contributions:

\begin{enumerate}
\item 
The first factor reflects the closeness of  two 
dijet invariant masses, $M_{jj}^{(1)}$ and $M_{jj}^{(2)}$,
to the nominal $W$ mass, $M_W$: 
$$
W_1 = W_a \,  W_b,  \qquad
W_a=      \mid M_{jj}^{(1)} - M_{jj}^{(2)}\mid  / \overline{M}_{jj}, \qquad
W_b =  \vert \overline{M}_{jj} - M_W \vert , 
$$
where $\overline{M}_{jj} = (M_{jj}^{(1)} + M_{jj}^{(2)}) /2$ represents
the average invariant mass of two dijets. The factor $W_a$ gets
small when there are two dijets with similar invariant masses, while $W_b$
is reduced when the average  mass of the  two dijets is close to the
nominal $W$ mass;

\item
If there are two dijets with the masses in the range  $M_W\pm 10\gev$,
these dijets have to be combined with the rest of the jets. This should
lead to several trijets which can be characterised by the invariant masses
$M_{jjj}$. For the top production, it is expected that there are at least two
trijets with similar invariant masses, $M_{jjj}^{(1)}$ and $M_{jjj}^{(2)}$. 
Therefore, one can introduce another factor:
$$
W_2 = \mid M_{jjj}^{(1)} - M_{jjj}^{(2)}\mid  / \overline{M}_{jjj},
$$
where $\overline{M}_{jjj} = (M_{jjj}^{(1)} + M_{jjj}^{(2)}) /2$ represents
the average invariant mass of two trijets.

\end{enumerate}

Each $k$-means cluster configuration
can be characterised by the factor $D_i=S_i \,  W_{1,i} \, W_{2,i}$
(the new index $i$ in $W_{1,i}$ and $W_{2,i}$ denotes a cluster
configuration obtained using a  certain  initial position of the centroids).
Only configurations with the smallest $D_i$ were accepted.
Since the clustering procedure minimizes $D_i$, rather than $S_i$,
the resulting particle assignment is the most optimal  
not only from the point of view of how well
hadrons are collimated in jets, but also how well such cluster configuration 
reflects the expected  $t\bar{t}$ decay property.
  
\begin{figure}
\begin{center}
\mbox{\epsfig{file=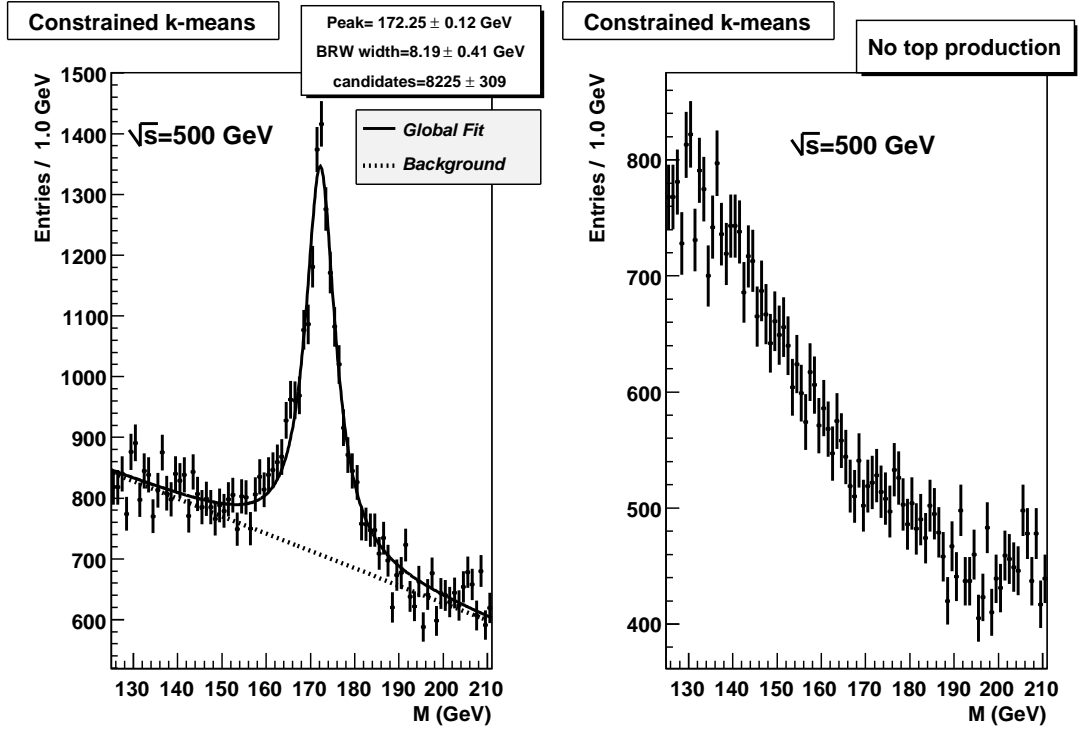,height=10cm}}
\caption
{
The dijet invariant masses for the all-hadronic top-decay
channel. Fully inclusive  $e^+e^-$ events were
generated with PYTHIA for $\sqrt{s}=500\gev$.
The reconstruction was done using the constrained $k$-means
algorithm (left). 
The fit was performed using the  Breit-Wigner 
function together with a second-order
polynomial to describe the background.
The invariant masses reconstructed with the
same algorithm using events without  $t\bar{t}$ production does not have 
a spurious peak near the nominal top mass (right plot).  
}
\label{ttfig2}
\end{center}
\end{figure}

The result of the constrained $k$-means algorithm is shown
in Fig.~\ref{ttfig2}(left). While the mass resolution and the systematic off-set
of the peak position are rather similar to the unconstrained version
of the algorithm, the efficiency of the constrained algorithm is significantly higher. 
Fig.~\ref{ttfig2}(right) shows the invariant masses for the background 
events (which do not contain the top events). 
The latter invariant mass does not show any structure near $175\gev$, 
indicating that the algorithm does not produce a spurious peak near $175\gev$.

Although we do not think that 
the computational speed is an important issue at the stage when no a detector simulation is
involved, a few words about the performance speed of the proposed algorithm
is still necessary. The (constrained) $k$-means jet algorithm is a factor two slower
than the Durham jet finder. However, the $k$-means algorithm requires
an additional pre-clustering stage for which the computational speed is rather similar  to 
that for the reconstruction of six jets by the Durham jet algorithm\footnote{All
the discussed jet algorithms were implemented  in C/C++.}.
Thus, the $k$-means procedure is roughly three times slower than the Durham algorithm.
Without the pre-clustering stage, the $k$-means algorithm is a factor 20--30 slower than the
Durham algorithm for the reconstruction of six jets.

\section{$W^+W^-$ production}

As a second example, let us consider
$e^+e^- \to W^+W^- \to 4 \;\hbox{\rm jets}$ at $\sqrt{s}=500\gev$. 
10k events were generated with PYTHIA using the same parameters
and the selection as before.
The $W$ mass was set to $80.45\gev$ and its width to $2.07\gev$.
We reconstructed  exactly four jets and then plot
the invariant masses of all six  jet pairs.
The $k$-means algorithm was constrained by the simple criteria: 
$D_i=S_i W_{1,i}$, where $W_1=\mid M_{jj}^{(1)} - M_{jj}^{(2)} \mid / \overline{M}_{jj}$
for each $k$-means clustering.

\begin{figure}
\begin{center}
\mbox{\epsfig{file=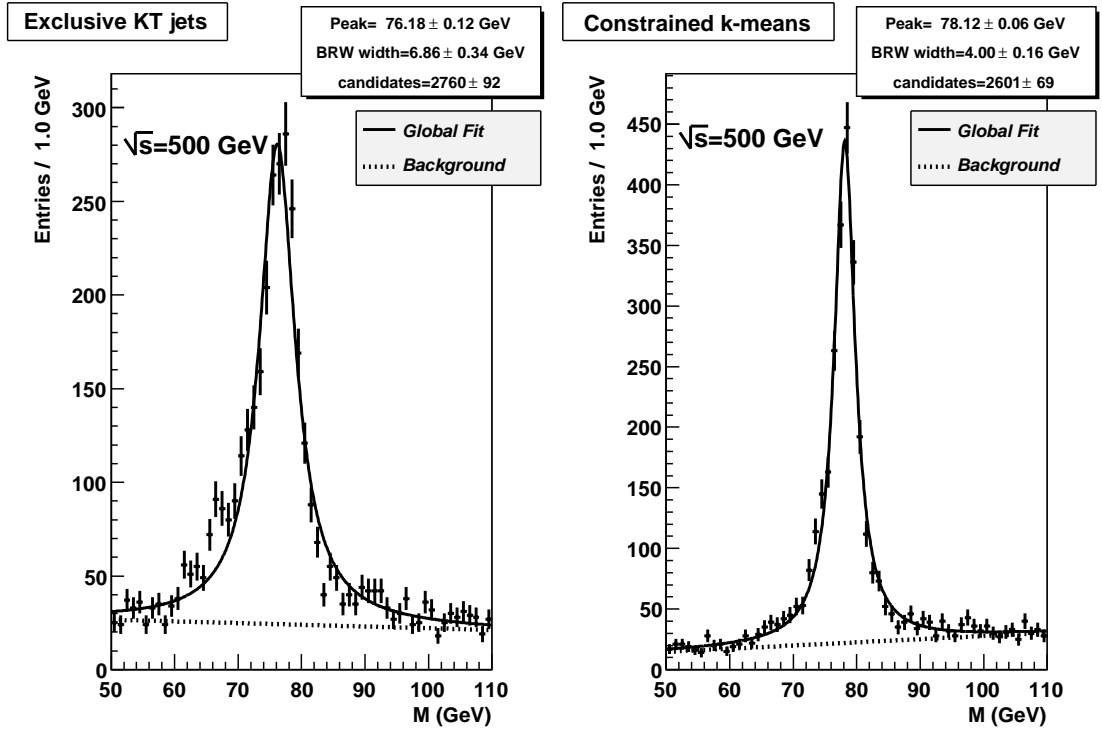,height=10cm}}
\caption
{
The dijet reconstructed invariant masses for the all-hadronic $W$-decay
channel  $e^{+}e^{-}\to W^+W^-\to 4\, \mathrm{jets}$. 
The events containing fully hadronic $W^+W^-$ decays were
generated with PYTHIA for $\sqrt{s}=500\gev$.
The reconstruction was done using the Durham algorithm
(left) and the constrained $k$-means algorithm (right).
The fit was performed using the  Breit-Wigner function together with a second-order
polynomial to describe the background.
}
\label{wwfig0}
\end{center}
\end{figure}

The results of the calculations are shown in Fig.~\ref{wwfig0}.
As before, the performance of the $k$-means algorithm is superior over the
Durham jet finder, especially for the reconstructed width. 
One may note that the Breit-Wigner peak shown in Fig.~\ref{wwfig0}(right) 
is also narrower than 
that for the invariant masses reconstructed with other traditional jet-finding 
algorithms \cite{Moretti:1998qx}.  
In addition, the systematical shift of 
the peak position reconstructed with the $k$-means procedure 
is smaller than for the Durham algorithm. However, the number of the reconstructed $W$ candidates
is somewhat  smaller than for the Durham algorithm.

\section{Conclusion}

A new jet clustering algorithm for the reconstruction of the   
invariant masses of heavy states decaying to hadronic jets 
was proposed\footnote{The C/C++ code of the constrained $k$-means algorithm   
is available as a module ``kmeansjets.rmc'' of the RunMC package \cite{runmc}.}. 
It is based on the $k$-means clustering procedure constrained by
additional kinematic requirements.

In this paper we did not try to cover many issues related to
the use of this algorithm. For example, we did not study    
the question of how to apply  this algorithm when no
fixed number of jets are expected, how to use this algorithm
in theoretical calculations, is this algorithm 
reliable in treating fixed-order perturbative QCD corrections and non-perturbative effects
and,  finally,  will a realistic event reconstruction with all detector effects included  
benefit from the use of this algorithm.  
All such issues have to be addressed in future.

Note that the constrained $k$-means clustering has nothing to do with the constrained fits
used in the invariant-mass reconstruction:  
The constrained fit attempts to
find the most optimal configuration when the error matrix on the
measured quantities are specified. The present  approach does not 
require such input  and it does not address  the issue 
of the experimental precision on the reconstructed jet energies and their
positions. Obviously, the constrained fit could also be used to 
improve the reconstruction of  heavy states from jet invariant masses.

For the proposed jet clustering,  
{\em a priory} specified 
physics requirements on event kinematics
can become an essential part of the minimisation procedure.
In contrast, the standard algorithms usually minimise a single distance measure.
The proposed algorithm has good reconstruction 
efficiency  and leads to a 
significantly better resolution for the invariant-mass reconstruction
than the traditional Durham jet finder. 
It is also expected that the peak
positions measured with the new algorithm
have small systematical uncertainty.
Finally, the proposed $k$-means approach can be used without
any physics constrain (which only increases the reconstructed efficiency), especially 
when the main issue is a good resolution on the
invariant-mass reconstruction.

\section*{Acknowledgements}
\vspace{0.3cm}

I thank V.~Morgunov who pointed out the importance of the particle  miss-assignment
problem for the top production at ILC. I also thank 
E.~Lohrmann for the discussion of the paper.

\newpage 

\bibliographystyle{./l4z_pl}
\def\bibname{\Large\bf References}
\def\refname{\Large\bf References}
\pagestyle{plain}
\bibliography{biblio}

\end{document}